\documentclass[twoside]{hepexdis07}
\usepackage[latin1]{inputenc}
\usepackage[dvips]{graphicx,epsfig,color}
\usepackage{wrapfig,rotating}
\usepackage{amssymb,amsmath,array}

\begin{document}
\title{Determination of $\Delta G\,/\,G$ from Open Charm events at COMPASS}

%***********************************************************************
% AUTHORS INFORMATION AREA
%***********************************************************************
\author{S. Koblitz
%
% Optional short acknowledgment: remove next line if non-needed
\thanks{supported by BMBF.}
, on behalf of the COMPASS Collaboration
%
% DO NOT MODIFY THE FOLLOWING '\vspace' ARGUMENT
\vspace{.3cm}\\
%
% Addresses and institutions (remove "1- " in case of a single institution)
Universit\"at Mainz - Institut f\"ur Kernphysik \\
Becherweg 45, 55099 Mainz - Germany\\
\phantom{w}\\
{\em  Talk given at the XV International Workshop on Deep-Inelastic}\\
{\em Scattering and Related Subjects, April 16-20, 2007, Munich, Germany }\\
http://www.mppmu.mpg.de/dis2007\\
}

%***********************************************************************
% END OF AUTHORS INFORMATION AREA
%***********************************************************************

\maketitle

\begin{abstract}
One of the main goals of the COMPASS experiment at CERN is the
determination of the gluon polarisation in the nucleon, $\Delta
G\,/\,G$. It is determined from spin asymmetries in the scattering of
$160\,{\rm GeV/}c$ polarised muons on a polarised LiD target.
The gluon polarisation is accessed by the selection of photon-gluon fusion
(PGF) events. A very clean selection of PGF events can be obtained with charmed
mesons in the final state.
Their detection is based on the reconstruction of $D^{\,\star}$ and
$D^{\,0}$ mesons in the COMPASS spectrometer.
The analysis method for the first measurement of $\Delta G/G$ from
the open charm channel is described. The result from COMPASS for the 2002-2004 data
taking period is shown. 
\end{abstract}

\section{Introduction}

In the framework of QCD, the spin of the nucleon is composed of the
contributions from the quark spin, $\Delta\Sigma$, and the gluon spin, $\Delta G$,
as well as the orbital angular momenta of quarks, $L_{q}$, and gluons, $L_{g}$:
$S_{N}=\frac{1}{2}=\frac{1}{2}\Delta\Sigma\,+\,\Delta
G\,+\,L_{q}\,+\,L_{g}$.
The discovery, that the quark contribution $\Delta\Sigma$ is
small \cite{EMC}, led to a series of measurements to determine the
other spin contributions.
Since QCD fits only give weak constraints on $\Delta G$, it has to be
measured directly. First investigations were performed by the
HERMES \cite{HERMES} and the SMC\cite{SMC} collaborations using high
$p_{t}$ hadron pairs in the final state. 
The primary goal of the COMPASS experiment is to perform
a precise measurement of $\frac{\Delta G}{G}$\, also with a new approach. 
Therefore charmed meson production is studied, since the
selection of charmed mesons in the final state provides an event
sample of photon-gluon fusion (PGF) events with no background from other
physical processes.

\section{$D$-Meson reconstruction}

\begin{wrapfigure}{l}{0.5\columnwidth}
\setlength{\unitlength}{1cm}
  \begin{picture}(10,6.5)
    \put(-0.7,-0.3){\includegraphics[width=7.8cm]{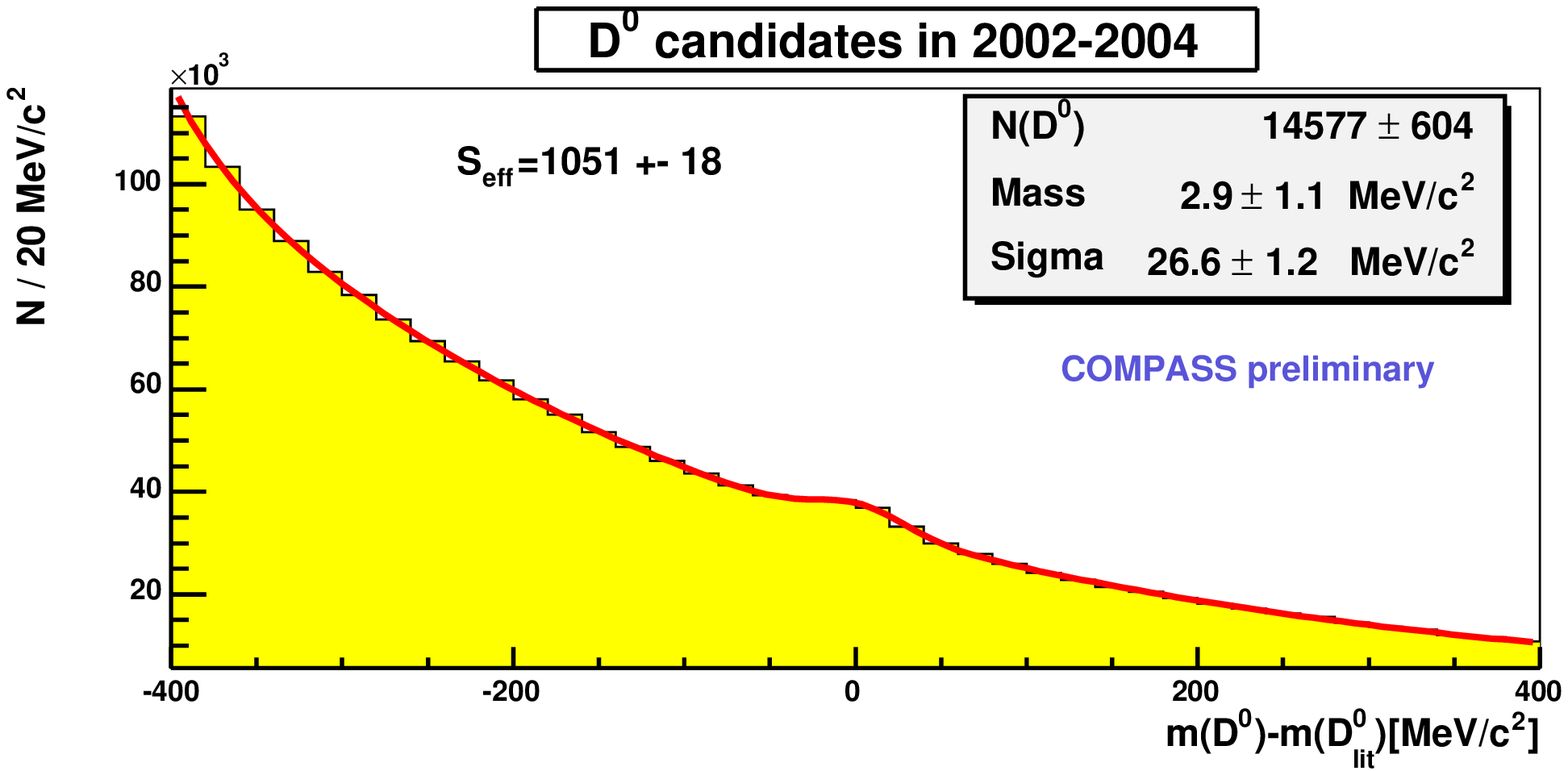}}
    \put(1.8,3.1){\includegraphics[width=6cm]{koblitz_susanne.fig4.eps}}
    \put(-0.7,3.1){\includegraphics[width=7.8cm]{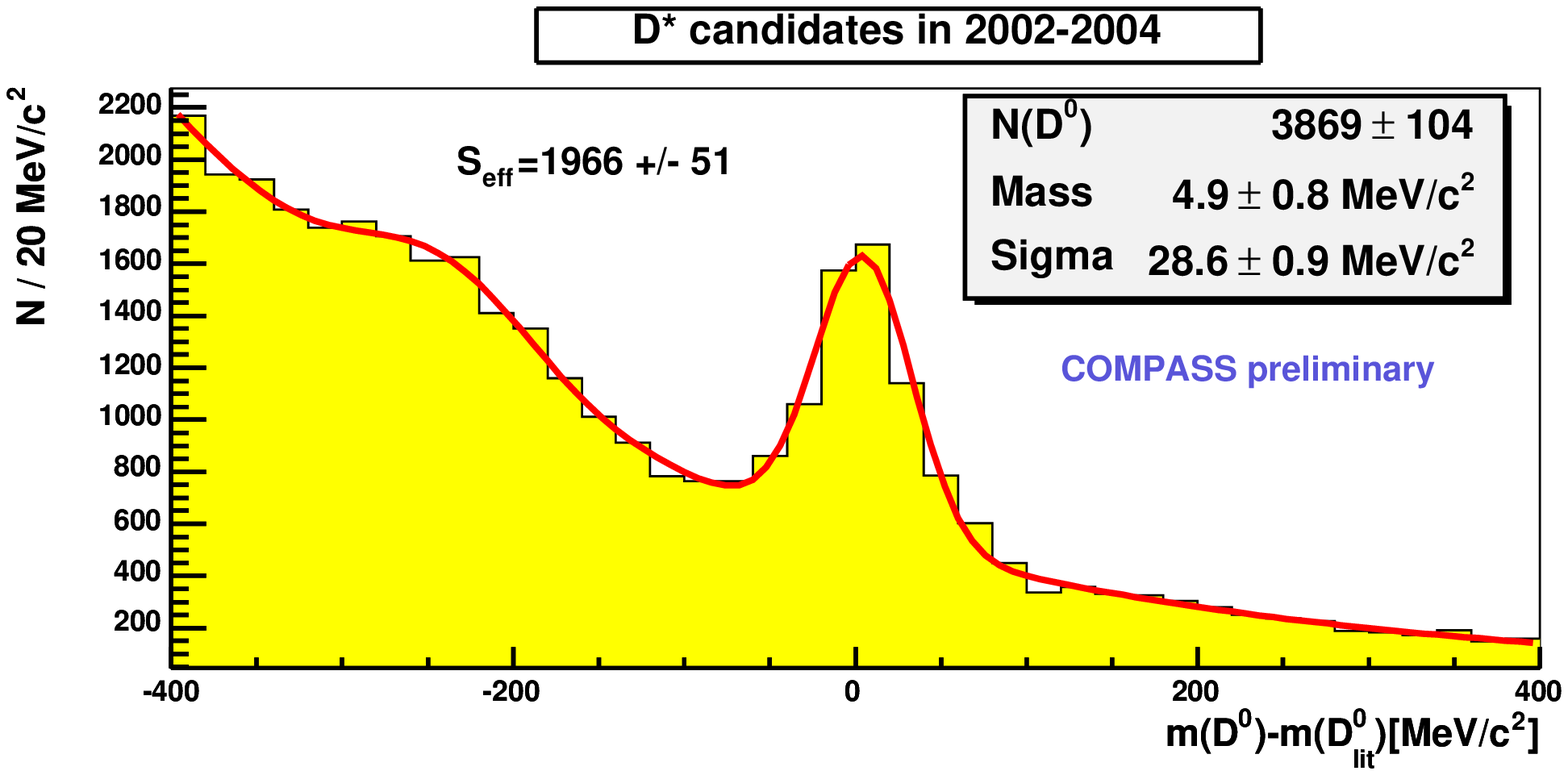}}
    \put(1.8,6.5){\includegraphics[width=6cm]{koblitz_susanne.fig4.eps}}
  \end{picture}
\caption{K$\pi$ mass spectra for \,$D^{\,\star}$\ tagged events (upper plot) and
the sample without \,$D^{\,\star}$\ tag ($S_{eff}\,=\,S^2\,/\,(B+S)\;$ is the
effective signal)}
\label{fig:dssig}
\vspace{-1.3cm}
\end{wrapfigure}

The PGF process is the main reaction for the production of charm
quarks in DIS. Due to the high charm mass, the charm content
of the nucleon can be neglected as well as the production of charm
quarks during fragmentation.
In the independent fragmentation of a $c\bar{c}$ pair most frequently
$D$~mesons are produced. On average 1.2 \,$D^{\,0}$\ mesons are produced per
each $c\bar{c}$ pair \cite{baum}.\\
The \,$D^{\,0}$\ mesons are reconstructed from their $K\,\pi$ decay which has
a branching ratio of $3.8\,\%$. The reconstruction is done using
tracks reconstructed in the COMPASS
spectrometer. A detailed description of the spectrometer can be found in
\cite{spectro}.
The thick nucleon target of the COMPASS experiment does not allow a
separation of production and decay vertex of the charmed meson.
Thus, the reconstruction of $D$ mesons is done on a combinatorial basis. For each
oppositely charged track pair in a given event the invariant mass is
calculated using
\begin{wrapfigure}{l}{0.5\columnwidth}
\setlength{\unitlength}{1cm}
  \begin{picture}(6,5.6)
    \put(0.2,-0.5){\includegraphics[width=6.5cm]{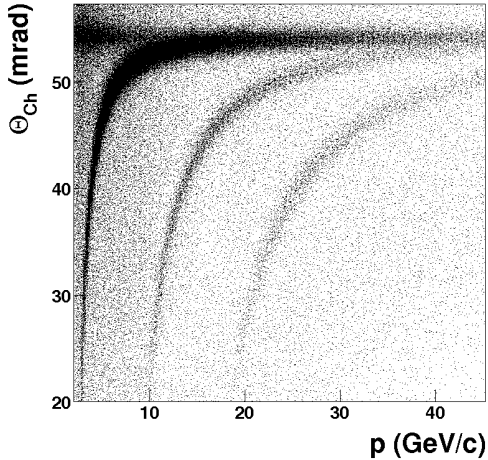}}
  \end{picture}
\caption{Cherenkov angles measured with RICH-1 vs. particle momenta. The histogram
content in the kaon region is multiplied by
a factor 30, in the proton region by a factor 150.}
\label{fig:RICH}
\vspace{-0.4cm}
\end{wrapfigure}
the kaon mass hypothesis for one of the tracks.\\
To suppress the high combinatorial
background several cuts are applied on the track pair. 
The most important requirement is the particle identification for the
kaon candidate from the Ring Imaging CHerenkov detector. The RICH allows to
separate $\pi$, $K$ and $p$ in a momentum range from the particle's
Cherenkov threshold to about $50\,{\rm GeV}$.
Figure~\ref{fig:RICH} shows
the RICH response for these three particle types as a function of their
momenta. As can be seen, kaons can be identified starting from the
kaon threshold around $9\,{\rm GeV}$.\\
Due to the large charm mass the fraction of energy from the virtual
photon that is carried by the meson, $z$, is expected to be higher for a
real charmed meson than for combinatorial background. Therefore a cut
of $z\,>\,0.25$ is applied on the \,$D^{\,0}$\ candidates. A third
cut to reduce the combinatorial background is applied
on the angle between the \,$D^{\,0}$\ flight direction and the $K$ momentum
vector in the \,$D^{\,0}$\ rest frame,
$|\cos{\theta_{K}^{\,\star}}|\,<\,0.5$.\\
With these cuts the ratio of open charm events to combinatorial
background is still in the order of
$1\,:\,10$ (cf. Fig.~\ref{fig:dssig}). Therefore, a second
more exclusive channel is also studied: $D^{\,\star}\,\rightarrow\,
D^{\,0}\,\pi\,\rightarrow \,K\,\pi\,\pi$. Due to the small mass
difference between \,$D^{\,\star}$\ and \,$D^{\,0}$, combinatorial background can be very
much suppressed by a cut on the mass difference: $3.1\,{\rm
MeV}\,<\, M_{K\pi\pi}\, -\,M_{K\pi}\, -\, M_{\pi}\, <\,9.1\,{\rm
MeV}$. Here, $M_{K\pi\pi}$ denotes the mass of the \,$D^{\,\star}$\ candidate and
$M_{K\pi}$ the mass of the \,$D^{\,0}$\ candidate. Since this so-called \,$D^{\,\star}$\
tag is very effective in the reduction of combinatorial background, the
$z$ and the $\cos{\theta^{\,\star}}$ cuts can be relaxed.
With
$z\,>\,0.2$ and $|\cos{\theta^{\,\star}}|\,<\,0.85$ for \,$D^{\,\star}$\ tagged
\,$D^{\,0}$\ mesons a signal to background ratio of 1:1 can be obtained
(cf. Fig.~\ref{fig:dssig}).

\section{Analysing Power}

The $\Delta G$\, measurement at COMPASS is based on the Photon Gluon Fusion
(PGF) process. In this process the photon emitted by the incoming muon
interacts with a gluon embedded in the nucleon. The interaction occurs
via the exchange of a virtual quark resulting in a $q\bar{q}$ pair in
the final state.
Studying the scattering of a polarised
muon beam off a polarised target gives access to experimental
muon-nucleon asymmetries of the tagged PGF process. 
To access the gluon polarisation $\Delta G$ information about the
hard subprocess is needed, which is combined in the analysing power
$a_{LL}$.
It contains the information about the partonic asymmetries from the
muon-gluon scattering process.
To determine $a_{LL}$ the kinematic variables of the hard subprocess
are needed. \\
Since only one of the two mesons is reconstructed, the full kinematics
of the PGF process is not known for each single event. 
Thus, a parametrisation based on measured quantities was introduced,
providing an estimation of
$a_{LL}$ for each open charm event. The parametrisation was obtained
by training a neural network with an event sample generated with the
AROMA generator in leading order QCD. For these events the full PGF
kinematics were 
available as well as the reconstructed observables
from the \,$D^{\,0}$\ mesons. The correlation between the $a_{LL}$ values
coming directly from the generated quantities and the reconstructed
$a_{LL}$ from the parametrisation is about $82\%$.
This procedure allows an evaluation of the analysing power for every
event entering the $\Delta G/G$ determination.

\section{Analysis Method}

The data analysis leading to $\Delta G/G$ is based on event rates for
scattering from a polarised muon beam off a polarised 
target. Since a separation between the
remaining background events and the signal events in the final event
sample is not possible, the
signal purity of the event sample, $\sigma_{PGF}/\sigma_{PGF}+
\sigma_{B}$, has to be introduced. It is determined from a fit to the
final mass spectra. To optimise the description of the signal purity,
this fit is done separately for events from the 
two target cells and for different bins of $a_{LL}$.\\
The observed event counting rates $N_{u,d}$ in the two
oppositely polarised target cells of the COMPASS target are related to
$\Delta G/G$ by
$$ N_{u,d} =  a\,\Phi\,n\,(\sigma_{PGF}+ \sigma_{B})(1+P_{{\rm
T}}\,P_{{\rm B}}\,f (a_{LL}\,\frac{\sigma_{PGF}}{\sigma_{PGF}+ \sigma_{B}}\,\frac{\Delta G}{G}
	+a_{LL}^B\,\frac{\sigma_{B}}{\sigma_{PGF}+\sigma_{B}}\,A_{B}))\;\;,$$
where $P_{{\rm B}}$ ($P_{{\rm T}}$) denotes the beam (target)
polarisation and $n$ the number of nuclei in the target. The dilution
factor $f$ describes the fraction of polarisable material in the
target. For ${}^{6} LiD$ this dilution factor is about $50\%$. 
The beam particle is required
to cross both target cells providing a cancellation of the beam 
flux $\Phi$. To cancel out the acceptance difference for the two cells, the
target spin orientation is reversed every eight hours, leading to a
total of 4 counting rates. From their double
ratio $$\delta=\frac{N_u\cdot N_d'}{N_u'\cdot N_d}\;\;$$
$\Delta G/G$ can be determined assuming a negligible background
asymmetry $A_{B}$ and a stable detector performance leading to a
cancellation of the acceptance factors, $$\frac{a_u\cdot
a_d'}{a_d\cdot a_u'} =1\;\;.$$
This method is applied to the events from every data taking period
as well as for the two channels separately. To improve the statistical
significance of the result, event weighting is used.
The final result is then
calculated as the weighted mean of the results for each channel and
data taking period.

\section{Results}

\begin{wrapfigure}{r}{0.55\columnwidth}
\setlength{\unitlength}{1cm}
  \begin{picture}(8,3.45)
    \put(-0.0,-0.4){\includegraphics[width=8cm]{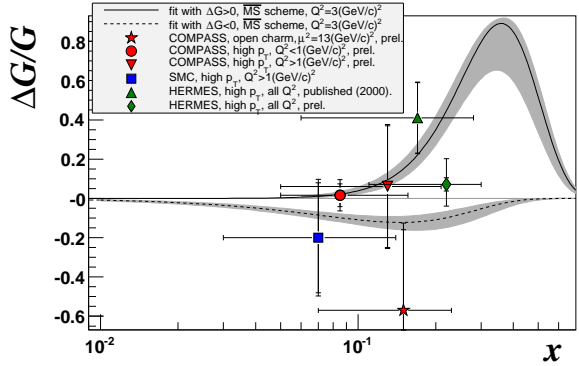}}
  \end{picture}
\caption{Comparison of the $\frac{\Delta G}{G}$\, measurements from COMPASS,
SMC~\cite{SMC} and HERMES~\cite{HERMES}. The curves show the
parametrisations at $3\,{\rm GeV}^2$ in the
$\overline{{\rm MS}}$ scheme from \cite{QCD}
Note that the open charm point was obtained at a
much higher scale.
}
\label{fig:dgg}
\vspace{-0.5cm}
\end{wrapfigure}
With the 2002-2004 data the preliminary result for the COMPASS open
charm analysis of 
$$<\frac{\Delta G}{G}>\,=\,-0.57\,\pm\,0.41({\rm stat})\,\pm\,0.17$$
was obtained. For the measured sample the average $x_g$ is $0.15$ with
RMS 0.08 and the hard scale at which this result was obtained is
$13\,{\rm GeV}^2$. \\
The largest contributions to the systematic uncertainty of this result
are possible false asymmetries (0.10), the choice of the fit function
for the signal purity (0.09) and possible background asymmetries
(0.07). There was no observation of any background asymmetry or false
asymmetry from detector instabilities, so the actual values of these
contributions are dominated by the statistical precision of the study. 
The influence of the choice for the fit function to describe the
signal purity was estimated using different fit functions in the
determination of $\Delta G/G$.\\
Further contributions to the
systematic uncertainty are coming from the choice of Monte Carlo
parameters (0.05), the number of bins for the signal-purity fit (0.04)
and the uncertainties of the dilution factor (0.03) and the polarisation
measurements (both 0.03). 

\section{Summary}

The result for the first $\frac{\Delta G}{G}$\, measurement from the open charm
channel is presented. This is the most direct measurement of $\frac{\Delta G}{G}$\,
since it is only weakly dependent on Monte Carlo simulation. 
A comparison of the COMPASS results and other existing results is
given in figure~\ref{fig:dgg}. The measurements are compared with the 
parton parametrisations from \cite{QCD}. The data points give an
indication that curves corresponding to small values of $\Delta G$\, are
favored. The analysis of the data taken in 2006 is in progress.


\begin{thebibliography}{99}
% Please replace the numbers for   contribId   and   sessionId
% in the following URL. You can get this information by going to 
% http://indico.cern.ch/confAuthorIndex.py?confId=9499
% and search for your contribution and click on the title
% Be aware: '&amp;' must be replaced by simple '&' as in example below
\bibitem{EMC}EMC, J. Ashman et al., {\em Phys. Lett.} {\bf B206}, 364 (1988).
\bibitem{HERMES}HERMES, A.Airapetian et al., {\em Phys. Rev. Lett.} {\bf
84}(2000) 2584. 
\bibitem{SMC}SMC, B. Adeva et al., {\em Phys. Rev.} {\bf D70} (2004)012002
%\bibitem{H1} C. Adloff et al., {\em Nucl.Phys.} {\bf B545} (1999) 21.
\bibitem{baum}COMPASS proposal, G. Baum et al., CERN/SPSLC 96-14.
\bibitem{spectro}COMPASS, P. Abbon et al., CERN-PH-EP/2007-001,
hep-ex/0703049, to be published in {\em Nucl. Inst. and Meth.}.
\bibitem{QCD}COMPASS, E.V.Yu. Alexakhin et al., {\em Phys. Lett} {\bf B647} 8 (2007).
\end{thebibliography}
\end{document}